\documentclass[twocolumn,english,aps,superscriptaddress]{revtex4}
\usepackage[T1]{fontenc}
\usepackage[latin1]{inputenc}
\usepackage{graphicx}

\makeatletter

\newcommand{\UCLA}{University of California at Los Angeles, Los
  Angeles, CA 90095--1547} 
\newcommand{\ASU}{Arizona State University, Tempe, Arizona
  85287--1504} 
\newcommand{\Saclay}{CEA-Saclay, Service de Physique Nucleaire,
  DAPNIA-SPhN, Cedex, France} 
\newcommand{\CMU}{Carnegie Mellon University, Pittsburgh, Pennslyvania
  15213} 
\newcommand{\CUA}{Catholic University of America, Washington,
  D.C. 20064} 
\newcommand{\CNU}{Christopher Newport University, Newport News,
  Virginia 23606} 
\newcommand{\UConn}{University of Connecticut, Storrs, Connecticut
  06269} 
\newcommand{\Duke}{Duke University, Durham, North Carolina
  27708--0305} 
\newcommand{\Edinburgh}{Edinburgh University, Edinburgh EH9 3JZ,
  United Kingdom} 
\newcommand{\FIU}{Florida International University, Miami, Florida
  33199} 
\newcommand{\FSU}{Florida State University, Tallahassee, Florida 32306} 
\newcommand{\Giessen}{Physikalisches Institut der Universit\"at
  Giessen, 35392 Giessen, Germany} 
\newcommand{\GWU}{The George Washington University, Washington, DC
  20052} 
\newcommand{\Glasgow}{University of Glasgow, Glasgow G12 8QQ, United
  Kingdom} 
\newcommand{\ISU}{Idaho State University, Pocatello, ID 83209}
\newcommand{\Frascati}{INFN, Laboratori Nazionali di Frascati,
  Frascati, Italy} 
\newcommand{\Genova}{INFN, Sezione di Genova, 16146 Genova, Italy} 
\newcommand{\Orsay}{Institut de Physique Nucleaire ORSAY, Orsay,
  France} 
\newcommand{\ITEP}{Institute of Theoretical and Experimental Physics,
  Moscow, 117259, Russia} 
\newcommand{\JMU}{James Madison University, Harrisonburg, Virginia
  22807} 
\newcommand{\KNU}{Kyungpook National University, Daegu 702-701, South
  Korea} 
\newcommand{\MIT}{Massachusetts Institute of Technology, Cambridge,
  Massachusetts 02139--4307} 
\newcommand{\UMass}{University of Massachusetts, Amherst,
  Massachusetts 01003} 
\newcommand{\UNH}{University of New Hampshire, Durham, New Hampshire
  03824--3568} 
\newcommand{\NSU}{Norfolk State University, Norfolk, Virginia 23504} 
\newcommand{\Ohio}{Ohio University, Athens, Ohio 45701} 
\newcommand{\ODU}{Old Dominion University, Norfolk, Virginia 23529} 
\newcommand{\PSU}{Penn State University, University Park, Pennsylvania
  16802} 
\newcommand{\Pitt}{University of Pittsburgh, Pittsburgh, Pennsylvania
  15260} 
\newcommand{\Roma}{Universita' di ROMA III, 00146 Roma, Italy} 
\newcommand{\RPI}{Rensselaer Polytechnic Institute, Troy, New York
  12180--3590} 
\newcommand{\Rice}{Rice University, Houston, Texas 77005--1892} 
\newcommand{\Richmond}{University of Richmond, Richmond, Virginia
  23173} 
\newcommand{\USC}{University of South Carolina, Columbia, South
  Carolina 29208} 
\newcommand{\JLab}{Thomas Jefferson National Accelerator Laboratory,
  Newport News, Virginia 23606} 
\newcommand{\Union}{Union College, Schenectady, NY 12308} 
\newcommand{\VPI}{Virginia Polytechnic Institute and State University,
  Blacksburg, Virginia 24061--0435} 
\newcommand{\UVa}{University of Virginia, Charlottesville, Virginia
  22901} 
\newcommand{\CWM}{College of Willliam and Mary, Williamsburg, Virginia
  23187--8795} 
\newcommand{\Yerevan}{Yerevan Physics Institute, 375036 Yerevan,
  Armenia} 

\usepackage{babel}
\makeatother
\begin{document}

\preprint{This line only printed with preprint option}

\title{Exclusive Photoproduction of the Cascade ($\Xi$) Hyperons}

\affiliation{\UCLA}
\affiliation{\ASU}
\affiliation{\Saclay}
\affiliation{\CMU}
\affiliation{\CUA}
\affiliation{\CNU}
\affiliation{\UConn}
\affiliation{\Duke}
\affiliation{\Edinburgh}
\affiliation{\FIU}
\affiliation{\FSU}
\affiliation{\Giessen}
\affiliation{\GWU}
\affiliation{\Glasgow}
\affiliation{\ISU}
\affiliation{\Frascati}
\affiliation{\Genova}
\affiliation{\Orsay}
\affiliation{\ITEP}
\affiliation{\JMU}
\affiliation{\KNU}
\affiliation{\MIT}
\affiliation{\UMass}
\affiliation{\UNH}
\affiliation{\NSU}
\affiliation{\Ohio}
\affiliation{\ODU}
\affiliation{\PSU}
\affiliation{\Pitt}
\affiliation{\Roma}
\affiliation{\RPI}
\affiliation{\Rice}
\affiliation{\Richmond}
\affiliation{\USC}
\affiliation{\JLab}
\affiliation{\Union}
\affiliation{\VPI}
\affiliation{\UVa}
\affiliation{\CWM}
\affiliation{\Yerevan}

\author{J. W. Price}
\email{price@physics.ucla.edu}
\affiliation{\UCLA}

\author{B. M. K. Nefkens}
\affiliation{\UCLA}

\author{J. L. Ducote}
\affiliation{\UCLA}

\author{J. T. Goetz}
\affiliation{\UCLA}

\author{G. Adams}
\affiliation{\RPI}

\author{P. Ambrozewicz}
\affiliation{\FIU}

\author{E. Anciant}
\affiliation{\Saclay}

\author{M. Anghinolfi}
\affiliation{\Genova}

\author{B. Asavapibhop}
\affiliation{\UMass}

\author{G. Audit}
\affiliation{\Saclay}

% ...

\collaboration{The CLAS Collaboration}
\noaffiliation

\date{\today}

\begin{abstract}
We report on the first measurement of exclusive $\Xi^-(1321)$ hyperon
photoproduction in $\gamma p\to K^+K^+\Xi^-$ for $3.2<E_\gamma<3.9$
GeV. The final state is identified by the missing mass in
$p(\gamma,K^+K^+)X$ measured with the CLAS detector at Jefferson
Laboratory.  We have detected a significant number of the ground-state
$\Xi^-(1321)\frac{1}{2}^+$, and have estimated the total cross section
for its production.  We have also observed the first excited state
$\Xi^-(1530)\frac{3}{2}^+$.  Photoproduction provides a copious source
of $\Xi$'s.  We discuss the possibilities of a search for the recently
proposed $\Xi_5^{--}$ and $\Xi_5^+$ pentaquarks. 
\end{abstract}

\maketitle

Little is known about the doubly-strange $\Xi$ hyperons. According to
the Review of Particle Properties (RPP), $J^P$ has been determined for
only three states: the $\Xi(1321)\frac{1}{2}^+$, the
$\Xi(1530)\frac{3}{2}^+$, and the
$\Xi(1820)\frac{3}{2}^-$~\cite{RPP04}. Eight more candidates have been
reported, but no $J^P$ determination has been made~\cite{RPP04}.
$SU(3)_F$ symmetry implies the existence of a $\Xi$ for every $N^*$
and also one for every $\Delta^*$~\cite{Nef95}. The RPP lists 24
well-established (3- or 4-star) $N^*$ and $\Delta^*$ resonances.
There are also 20 $N^*$ and $\Delta^*$ ``candidates'' (1- or 2-star).
We therefore expect to find at least 24 $\Xi^*$ states; another 20
states may also exist. 

Because the cascades have strangeness $S=-2$, they are difficult to
produce.  The study of these hyperons has thus far centered on their
production in $K^-p$ reactions; some $\Xi^*$ states were found using
high energy hyperon beams. It is important to find other means of
$\Xi$ production --- there is no suitable $K^-$ facility for the
production of the excited $\Xi^*$ states available now or in the
forseeable future. 

The inclusive photoproduction process $\gamma p\to\Xi^-X$ has been
studied by two groups.  In both cases, the $\Xi^-$ was reconstructed
from the decay products in the chain
$\Xi^-\to\pi^-\Lambda\to\pi^-\pi^-p$.  Aston \emph{et
  al.}~\cite{Ast82} used a tagged photon beam in the energy range
$20<E_\gamma<70$ GeV at the CERN SPS with the Omega spectrometer, and
measured a cross section of $28\pm9$ nb for
$x_F(=2p^*_\parallel/\sqrt{s})>-0.3$. Abe \emph{et al.}~\cite{Abe85}
used a 20 GeV laser-backscattered photon beam incident on the SLAC 1-m
hydrogen bubble chamber and quote a total cross section of
$117\pm17$ nb.  They also report a value of $94\pm13$ nb in the same
$x_F$ range as the CERN group, in strong disagreement with Aston
\emph{et al.}~\cite{Ast82}.  The discrepancy between these two
experiments has never been addressed.

The availablility at the Thomas Jefferson National Accelerator
Facility (JLab) of high-energy, high-quality photon and electron beams
up to 6 GeV suggests that the prospects for cascade photoproduction
should be revisited.  All 11 cascade states listed in the RPP are very
narrow ($9-60$ MeV)~\cite{RPP04}, and there is reason to believe that
any missing cascades are also narrow~\cite{Ris03}.  The $\Xi^-$ can
therefore be observed as a sharp peak in the missing mass spectrum in
$p(\gamma,K^+K^+)X$.  This method has a great benefit in that it can
be used without modification to search for all narrow excited cascade
states~\cite{Nef96}.

In this paper, we present the first measurement of exclusive $\Xi^-$
photoproduction in the process $\gamma p\to K^+K^+\Xi^-$.  We use the
missing mass technique to measure the cross section for the production
of the ground state $\Xi^-$, and establish a signal for the first
excited state $\Xi^-(1530)$.  The method, while currently limited by
statistics, is a viable option for future searches for high-mass
$\Xi^*$ states.  The availability of a substantial sample of cascade
hyperons, both in the ground state and excited states, will allow the
pursuit of several avenues of research~\cite{Pri02}.  These include
the search for the many missing cascade states mentioned above,
studies of interesting cascade decays, $J^P$ measurements of the $\Xi$
states, the \emph{s-d} quark mass difference, and with a long target
to allow rescattering, $\Xi p$ scattering and double $\Lambda$
hypernuclear production. 

The interest in cascade physics has received a major boost due to the
recent evidence for the production of pentaquarks, even though their
existence is not firmly established~\cite{Pen03,Pen04}.  Within the
proposed antidecuplet of pentaquark states, three are \emph{manifestly
exotic}, in that their quantum numbers preclude them from being
three-quark states: the $\Theta^+(1540)$, the $\Xi_5^{--}$, and the
$\Xi_5^+$ (the subscript ``5'' refers to the pentaquark nature of
these objects).  Only one experiment, NA49, has claimed a signal for
the $\Xi_5$~\cite{Alt04}, although there is dissention within the NA49
group as to the interpretation of this result~\cite{Fis04}.  Other
experiments with much
higherstatistics~\cite{Ada98,Kno04,CDF04,Che04,Aub04} have not seen
this state.  The RPP rates the $\Xi_5$ as a one-star
state~\cite{RPP04}.  It is urgently necessary to find a complementary
approach to investigate the existence of the $\Xi_5$.

The photon energy threshold for the production of the ground state
$\Xi^-(1321)\frac{1}{2}^+$ is 2.4 GeV; the first excited state, the
$\Xi^-(1530)\frac{3}{2}^+$, requires $E_\gamma>2.96$ GeV. These
energies are readily available at JLab with the Hall B Photon
Tagger~\cite{Sob00}, while the two $K^+$'s can be detected with the
large-acceptance multi-particle spectrometer CLAS~\cite{Mec03}.  This
detector is a six-sector spectrometer with a toroidal magnetic field.
Three sets of drift chambers surrounded by a highly-segmented
scintillation counter system determine the momentum and velocity of
the outgoing charged particles at polar angles in the range
$10^\circ-140^\circ$.

To establish that there are two $K^+$'s in the final state,
time-of-flight is used over a $\sim$5-m flight path to the outermost
layer of the CLAS detector.  This makes the detection efficiency
strongly dependent on the kaon momentum.  Each kaon must also have
enough perpendicular momentum to be directed into the detector, which
places a limit on the maximum observable $\Xi^*$ mass beyond that
imposed by the photon energy.  These factors are partially offset by
the toroidal field of the CLAS magnet, which bends positively-charged
particles away from the beamline.  The measurement described in this
paper benefits from the resulting large geometrical acceptance for the
two $K^+$'s. 

We have analyzed two existing CLAS data sets for the exclusive
photoproduction process $p(\gamma,K^+K^+)\Xi^-$.  The first data set,
labeled $g6a$, had a photon energy range $3.2<E_\gamma<3.9$ GeV, with
a photon flux of $10^6\,\gamma/s$.  For the second data set, $g6b$,
the photon energy range was $3.0<E_\gamma<5.2$ GeV, and the photon
flux was approximately five times higher.  The running conditions for
the two data sets were otherwise identical.  An 18-cm-long
liquid-hydrogen target was located at the center of CLAS.  The trigger
required a coincidence between charged particle tracks in two opposing
sectors. The integrated luminosity of the $g6a$ data set is
1.1~pb$^{-1}$.  The luminosity of the $g6b$ set is approximately twice
as large, but the absolute normalization uncertainties in this data
set prevent us from using it in our evaluation of the cross section.
The determination of the photon flux is discussed in~\cite{Sob00}. 

\begin{figure}
\centerline{\includegraphics[%
  width=1.0\columnwidth,
  keepaspectratio]{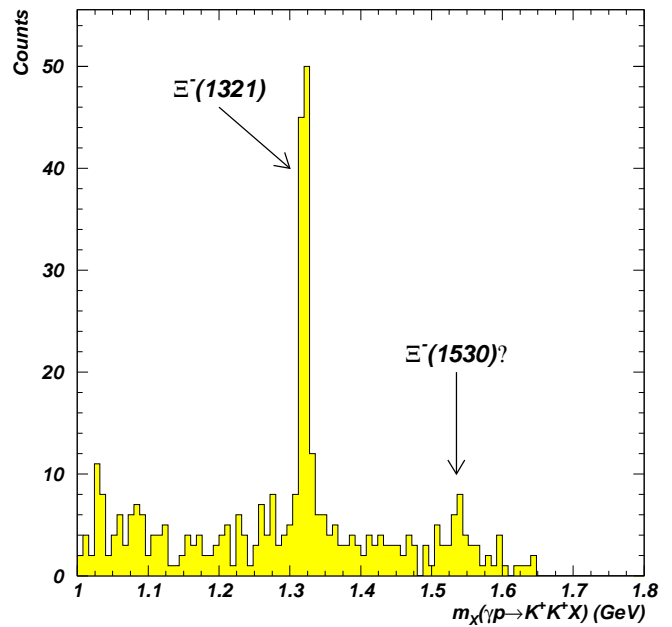}}
\caption{\label{fig:g6a-mmass}The missing mass $m_X$ in the process
  $p(\gamma,K^+K^+)X$ for the $g6a$ data set.  The figure has not been
  corrected for acceptance.  The ground-state
  $\Xi^-(1321)\frac{1}{2}^+$ is clearly seen; the signal-to-background
  ratio exceeds 10:1. A possible enhancement is seen in the plot at
  the position of the first excited state
  $\Xi^-(1530)\frac{3}{2}^+$. The arrow indicates the RPP mass of this
  state at 1.535 GeV.} 
\end{figure}

The identification of a particle as a $K^+$ is based on the measured
momentum and velocity.  Figure~\ref{fig:g6a-mmass} shows the missing
mass spectrum for the process $p(\gamma,K^+K^+)X$ from the $g6a$ data
set. 
This spectrum has not been corrected for acceptance.  There is a
large, very narrow peak at 1320 MeV, with a signal-to-background ratio
of better than 10:1.  The mass is in excellent agreement with the
RPP value of $1321.3\pm0.1$ MeV~\cite{RPP04}.  This excellent
agreement is likely fortuitous; based on other measured missing
masses in this analysis, we estimate the systematic uncertainty in the
mass determination of this data set is about 10 MeV.  Subtracting a
polynomial background, we find that the ground-state peak has
$101\pm12$ events.  A flat background yields nearly the same result.
The FWHM is approximately 15 MeV, consistent with the missing mass
resolution of the CLAS detector. 

\begin{figure}
\centerline{\includegraphics[%
  width=1.0\columnwidth,
  keepaspectratio]{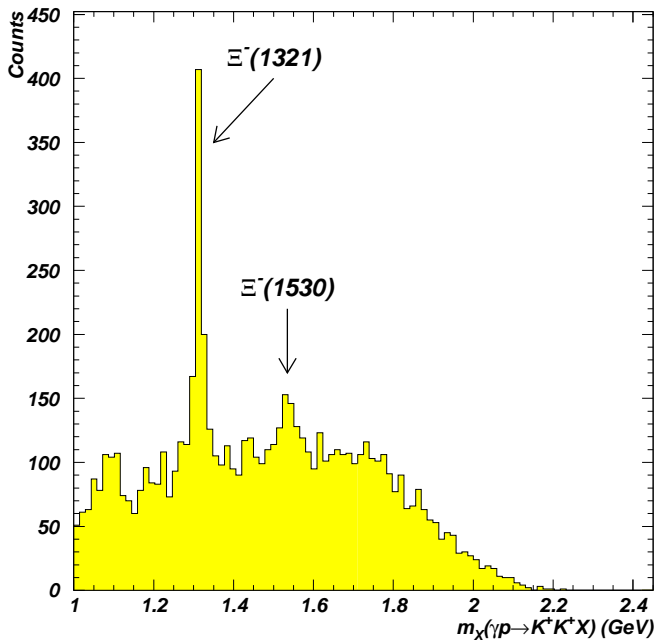}}
\caption{\label{fig:g6b-mmass}Same as Fig.~\ref{fig:g6a-mmass} for the
  $g6b$ data set.  Both the ground state $\Xi^-(1321)\frac{1}{2}^+$
  and the first excited state $\Xi^-(1530)\frac{3}{2}^+$ are seen.}
\end{figure}

There is an indication of a small peak in Fig.~\ref{fig:g6a-mmass} in
good agreement with the mass of the first excited state, the
$\Xi^-(1530)\frac{3}{2}^+$.  It is too small for any conclusion about
the $\Xi^-(1530)$ to be made, due to the relatively low available
photon energy and a reduced detector acceptance.  To investigate this,
we have analyzed the CLAS \emph{g6b} data set.
Figure~\ref{fig:g6b-mmass} shows the corresponding missing mass
spectrum. 
In this data set the $\Xi^-(1530)$ is clearly visible. Higher-mass
states, however, cannot be seen above the background. A data set
concentrated at high energy, the CLAS \emph{g6c} data set
($4.8<E_\gamma<5.4$ GeV; $\int\mathcal{L}dt=2.7\,\mathrm{pb}^{-1}$),
will be analyzed for the production of heavier excited states of the
cascade. This analysis will be the subject of a future publication.

One of the advantages of the missing mass technique is that the
physics backgrounds are small; if the final state contains two
$K^+$'s, whatever is left must have the quantum numbers of the
$\Xi^-$.  The first real background that can appear is due to the
process $\gamma p\to K^+\phi\Lambda$, where the $\phi$ decays to
$K^+K^-$.  This background only contributes for missing masses above
$m_{K^-}+m_\Lambda=1.6$~GeV.  

The high photon flux contributes to another background due to $K/\pi$
misidentification.  The analysis procedure used for this data matched
the timing of each track in CLAS to that of a tagged photon in our
photon tagger.  The innermost timing detector in CLAS had only three
channels, which resulted in a large accidental background from one of
two likely final states: $\pi^+\pi^+\Delta^-$ and $K^+\pi^+\Sigma^-$.
Both of these can appear to be the $K^+K^+\Xi^-$ final state if the
photon that caused the event was not tagged (if, for instance, it was
below the range of the photon tagger), while a higher-energy photon
was tagged nearby in time.  This results in the large background in
Fig.~\ref{fig:g6b-mmass}.

Even when the photon is tagged correctly, a high-energy pion can
masquerade as a high-energy kaon.  If this happens in the process
$\gamma p\to K^+\pi^+\Sigma^-$, the resulting missing mass will be
incorrectly calculated.  This results in the enhancement in
Fig.~\ref{fig:g6b-mmass} near 1100 MeV.  A similar background, due to
the process $\gamma p\to K^+\pi^+\Sigma^-(1385)$, is expected to
appear near the mass of the ground-state cascade.  

The large background under the peak in Fig.~\ref{fig:g6b-mmass}, along
with the $g6b$ normalization difficulty mentioned earlier, makes the
extraction of a cross section difficult.  We therefore do not report
cross sections for this data set, as improvements have been made to
the CLAS detector to mitigate both of these issues.  Future data
is expected to be much cleaner.  We may still use the $g6b$ data set
to make a qualitative assessment of the feasibility of the $\Xi$
photoproduction program. 

\begin{figure}
\centerline{\includegraphics[%
  width=1.0\columnwidth,
  keepaspectratio]{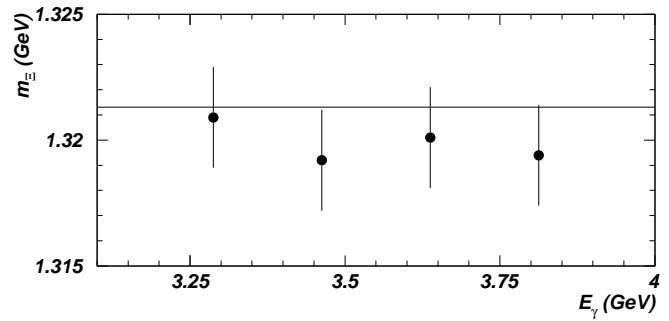}}
\caption{\label{fig:ximassconsistency}Consistency of the missing mass
  of the $K^+K^+$ system. Shown is the centroid of the peak in the
  $K^+K^+$ missing mass for $\gamma p\to K^+K^+X$ as a function of the
  incident photon energy. The vertical error represents one-half the
  bin width of the plot in Fig.~\ref{fig:g6a-mmass}. The horizontal
  line is the PDG value of the $\Xi^-$ mass of 1321.31 MeV.} 
\end{figure}

We can show that a peak is not an artifact of $K/\pi$
misidentification by investigating the dependence of the position of
the peak on the incident photon energy and the cascade production
angle. By dividing our data into four $E_\gamma$ bins, we effectively
make four independent measurements of the $\Xi^-$ mass.  As seen in
Fig.~\ref{fig:ximassconsistency}, the peak position is stable over a
700-MeV $E_\gamma$ range.
A similar test was performed, plotting the peak position as a function
of the $\Xi^-$ c.m.\ angle, with the same results.

The cascade production mechanism is not known at present; it likely
involves the intermediate production of any of several high-mass $N^*$
and $Y^*$ states. This makes the calculation of the acceptance
difficult and results in a large systematic uncertainty in the
extraction of the cross section.  Our estimate of the cross section is
based on a uniform $K^+K^+\Xi^-$ phase space distribution of the final
state particles. In this special case, we used a Monte Carlo
simulation of the CLAS detector based on GEANT 3.21 to find that the
ground state photoproduction process for the $g6a$ data set has an
acceptance of $2.8$\%, averaged over the entire $E_\gamma$ range.

The dominant systematic uncertainty is in the acceptance calculation,
due to the unknown production mechanism for this process.  The limited
statistics of this measurement prevent us from making a detailed study
of the production, but we may make an estimate of the effect of
different production models by comparing the acceptance based on our
phase-space calculation above with a toy model in which the cross
section varies as a function of the momentum transfer $t$ to the
$K^+K^+$ system, with the functional form $\sigma=Ae^{Bt}$.  For such
a model, we obtain a smaller acceptance and a correspondingly higher
cross section.  By comparing the simulation with the data, we obtain
$B=1\pm1$, leading to a variation in the calculated acceptance of
$\sim30\%$.  We use this as our systematic uncertainty, and obtain a
value for the total cross section, averaged over the photon energy
range $3.2<E_\gamma<3.9$ GeV, of $3.5\pm0.5(stat.)\pm1.0(syst.)$ nb.

At luminosites attainable with photon experiments with the CLAS
detector, our data imply the production of several thousand cascade
ground-state hyperons per week.  The cross section for the first
excited state is roughly two to three times smaller than for the
ground state, but nevertheless provides a reasonable counting rate
in a dedicated experiment.  We are therefore confident that we will
have sufficient counting rates to justify initiating the program of
cascade physics outlined in Ref.~\cite{Pri02}. 

With small modifications, the photoproduction method may be used to
search for $\Xi_5$ pentaquarks.  In the prediction of
Ref.~\cite{Dia97} and elsewhere, the $\Xi_5$ has isospin 3/2, with
$-2\le Q\le+1$.  The $\Xi_5^-$ can be detected using the process
$p(\gamma,K^+K^+)\Xi_5^-$, similar to the 3-quark $\Xi^-$.  To detect
the other three charge states, additional pions of the appropriate
charge can be added to the final state.  The processes
$p(\gamma,K^+K^+\pi^+)\Xi_5^{--}$ and
$p(\gamma,K^+K^+\pi^-\pi^-)\Xi_5^+$ would be used to detect the two
manifestly exotic cascades.  Because these processes have extra
particles in the final state, they also have correspondingly higher
photon energy thresholds.  It is therefore necessary to run at the
highest energies possible for these searches.  The identification of
the $\Xi_5^-$ and the $\Xi_5^0$ as pentaquarks is dependent on also
finding the $\Xi_5^{--}$ or the $\Xi_5^+$ at the same mass.  If the
pentaquarks are found, we may use the process
$p(\gamma,K^+K^+)\Xi_5^-$ to compare the properties of the pentaquark
cascades with those of the 3-quark cascades, such as mass splittings,
widths, decay rates, and decay modes.  The ability to look at both of
these types of states with the exact same process makes this a
powerful approach.

We have demonstrated the potential of photoproduction to investigate
new cascade states.  Furthermore, we have shown that JLab has
sufficient energy and tagged photon flux for this purpose. The
absolute energy calibration of the Hall B photon tagger and CLAS allow
the determination of missing masses to $<1\%$ in the cascade mass
region.  This method provides a complementary approach to the search
for the cascade pentaquark.

We would like to acknowledge the outstanding efforts of the staff of
the Accelerator and the Physics Divisions at JLab in support of this
experiment.  This work was supported in part by the U.S. Department of
Energy, the National Science Foundation, Emmy Noether grant from the
Deutsche Forschungs Gemeinschaft, the French Centre National de la
Recherche Scientifique, the French Commissariat \`a l'Energie
Atomique, the Istituto Nazionale di Fisica Nucleare, and the Korea
Research Foundation.  The Southeastern Universities Research
Association (SURA) operates the Thomas Jefferson National Accelerator
Facility for the United States Department of Energy under contract
DE-AC05-84ER40150.

\end{document}